\documentclass[sigconf,nonacm]{acmart}
\usepackage{algorithm}
\usepackage{utfsym}
\usepackage{algpseudocode}
\usepackage{multirow}
\usepackage{booktabs}
\usepackage{pifont}
\usepackage{tikz}
\usepackage{xifthen} 
\usepackage[dvipsnames, svgnames, x11names]{xcolor} 
\usepackage{graphicx}
\usepackage{subcaption}

\newcommand{\bangou}{\textcolor{ProcessBlue}{\ding{52}\rotatebox[origin=c]{-9.2}{\kern-0.7em\ding{55}}}}

\definecolor{cG}{rgb}{0.35,0.75,0.5}
\definecolor{cR}{rgb}{0.75,0.1,0.1}
\AtBeginDocument{%
  }

\usepackage{fontawesome}
\usepackage{cleveref}

\begin{document}

\newcommand{\copytmp}[1]{\textcolor{red}{#1}}
\newcommand{\notsure}[1]{\textcolor{blue}{#1}}
\newcommand{\todo}[1]{\textcolor{purple}{#1}}
\newcommand{\yes}[1]{\textcolor{cG}{\usym{1F5F8}} }
\newcommand{\no}[1]{\textcolor{cR}{\usym{2717}}}

\newcommand{\controlline}[3]{%
  \ifthenelse{\equal{#1}{1}}{#2}{#3}%
}

\title{Multi-Plane HyperX: A Low-Latency and Cost-Effective Network for Large-Scale AI and HPC Systems}


\author{Ziyu Wang}
\orcid{0009-0001-6074-1199}
\affiliation{%
  \institution{College of Computer Science and Technology, National University of Defense Technology}
  \city{Changsha}
  \state{Hunan}
  \country{China}
}
\email{wangziyu@nudt.edu.cn}

\author{Fei Lei}
\orcid{0000-0001-8614-933X}
\affiliation{%
  \institution{College of Computer Science and Technology, National University of Defense Technology}
  \city{Changsha}
  \state{Hunan}
  \country{China}
}
\email{leifei@nudt.edu.cn}

\author{Dezun Dong} 
\authornote{Corresponding author}
\orcid{0000-0001-6243-8479}
\affiliation{%
  \institution{College of Computer Science and Technology, National University of Defense Technology}
  \city{Changsha}
  \state{Hunan}
  \country{China}
}
\email{dong@nudt.edu.cn}

\renewcommand{\shortauthors}{Anonymous.}



\begin{abstract}
Multi-plane architectures have become increasingly prevalent in the Fat-Tree networks of AI data centers. By leveraging multiple ports on a single network interface card (NIC) or multiple NICs within a scale-up domain, each port or NIC is allocated to an independent network plane, thereby provisioning the overall system with multiple network planes. However, no prior literature has explored the application of multi-plane technologies to direct networks such as HyperX. This paper investigates the multi-plane HyperX network and demonstrates that, compared to state-of-the-art network topologies like multi-plane Fat-Tree, Dragonfly, and Dragonfly+, the multi-plane HyperX architecture achieves a significantly smaller network diameter and superior cost-effectiveness.

\end{abstract}


\maketitle

\section{INTRODUCTION}

Recently, various multi-plane Fat-Tree networks have been proposed in both academia and industry. For instance, Weiyang \textit{et al.} introduced the Rail-only \cite{rail_only24} network, where each Network Interface Card (NIC) within a high-bandwidth domain belongs to an independent Fat-Tree network plane. Alibaba proposed the HPN 7.0 \cite{hpn7_24} network, which significantly reduces the risk of hash collisions in Equal-Cost Multi-Path (ECMP) routing by leveraging dual-plane network characteristics. DeepSeek proposed an ideal multi-plane network \cite{deepseekv3_25} architecture where a NIC is equipped with multiple physical ports, each belonging to an independent Fat-Tree plane. Furthermore, the multi-plane network proposed by DeepSeek requires NICs to uniformly spray traffic across all physical ports; this necessitates that the NICs possess switching functionalities and provide native support for packet-spraying technologies and out-of-order packet reception.

The benefits of multi-plane Fat-Tree networks are substantial, as they eliminate the core layer of traditional three-tier Fat-Tree networks, thereby significantly reducing both network diameter and costs. However, current research on multi-plane technology remains predominantly limited to indirect topologies like Fat-Tree. Direct topologies, such as Dragonfly \cite{Dragonfly08}, Flattened Butterfly \cite{flattened07}, and HyperX \cite{HyperX09}, are widely deployed in leadership-class large-scale HPC systems and supercomputers. In this paper, we pioneer the application of multi-plane technology to the HyperX direct topology. HyperX is a superset of the Flattened Butterfly network and comprises multiple dimensions, where switches within each dimension are interconnected in a full-mesh configuration. We demonstrate that the multi-plane HyperX network achieves smaller network diameter and superior cost-effectiveness compared to several state-of-the-art topologies, including multi-plane Fat-Tree, Dragonfly, and Dragonfly+.

\section{MOTIVATION}

Driven by considerations such as fault tolerance, Network Interface Cards (NICs) in many contemporary large-scale AI and HPC systems are equipped with multiple ports, establishing the foundation for constructing multi-plane networks. A multi-port NIC evenly distributes its total outbound bandwidth across all its physical ports. Concurrently, switches typically support various modes of port breakout; for example, the 51.2~Tbps switch deployed in Alibaba's HPN~7.0 supports breakout configurations of $128 \times 400$~Gbps and $256 \times 200$~Gbps. Consequently, it is a natural approach to simultaneously apply port breakouts to both the NICs and the switches, ensuring their bandwidths match. Performing port breakouts on a switch significantly increases its radix, which enables the reduction of the network diameter when constructing a network of the same scale, thereby achieving lower network latency and superior cost-effectiveness. Guided by this concept, we can construct the \textbf{M}ulti-\textbf{P}lane \textbf{H}yper\textbf{X} \textbf{(MPHX)} network, which will be detailed in the following section.

\section{DESIGN}

\begin{figure}[t]
    \centering
    \includegraphics[width=0.9\linewidth]{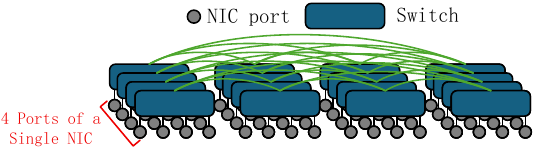}
    \caption{4-plane 1D HyperX (MPHX(4,4,4)) network. Each NIC is equipped with four ports, with each port belonging to an independent 1D HyperX network plane.}
    \label{1Dmphx}
\end{figure}

The HyperX network is a superset of the Flattened Butterfly network, featuring multiple dimensions where switches within each dimension are fully interconnected. The primary distinction between the HyperX and Flattened Butterfly topologies is that HyperX allows for a more flexible distribution of the number of switches across each dimension. For further information regarding the HyperX network, please refer to \cite{HyperX09}. Next, we detail the construction of the multi-plane HyperX network. Table \ref{symbols} summarizes the notation used in this paper. We define the total outbound bandwidth of a NIC as $B$ and the total bandwidth of a switch as $B \times k$, indicating that it can be configured with a maximum of $k$ ports, each with a bandwidth of $B$. The number of NIC ports (which corresponds to the number of independent network planes) is denoted by $n$, yielding a bandwidth of $\frac{B}{n}$ for each NIC port. Consequently, a switch can also be configured in a $\frac{B}{n'} \times n'k$ port breakout format (in default, we assume $n'=n$). Each switch connects to $p$ NIC ports. Assuming the HyperX network has $D$ dimensions, and letting $D_1, D_2, \dots, D_D$ represent the number of switches in each respective dimension, a multi-plane HyperX network can be denoted as MPHX$(n, p, D_1, D_1, \dots, D_D)$.

Equation~\ref{mphx_eq1} defines the total number of Network Interface Cards (NICs) that a HyperX network can accommodate:
\begin{equation}\label{mphx_eq1}
N = p \times \prod \limits_{i=1}^D D_i 
\end{equation}
In a balanced, maximum-scale multi-plane HyperX network, the configuration should satisfy $p=D_1=D_2=\dots=D_D=\frac{nk}{D+1}$. Consequently, Equation~\ref{mphx_eq2} presents the total number of NICs that can be connected within this maximum-scale architecture:
\begin{equation}\label{mphx_eq2}
N_{max} = \left(\frac{nk}{D+1}\right)^{D+1}
\end{equation}

\begin{table}[t] 
  \caption{Symbols used in this paper}
  \label{symbols}
  \centering
  \begin{tabular}{c|c}
    \toprule 
    Symbols & Explain \\
    \midrule
    $N$ & The number of NICs of the network\\
    $B$ & The total outbound bandwidth of a NIC\\
    $n$ & The number of ports of each NIC \\
    $B \times k$ & The bandwidth of each switch\\
    $D$ & The number of dimensions of the HyperX network\\
    $p$ & The number of connected NIC ports of each switch\\
    $d$ & The diameter of the network\\
    $N_s$ & The number of switches in the network\\
    $N_o$ & The number of optical modules in the network\\
    
  \bottomrule
\end{tabular}
\end{table}

In practice, the value of $n$ is bounded. In this paper, we assume the maximum value of $n$ to be 8, meaning that a single NIC can be broken out into at most 8 ports, resulting in a maximum of 8 network planes.

\section{COST}

In this section, we compare the cost-effectiveness of the three-tier Fat-Tree, multi-plane Fat-Tree, and MPHX networks. We assume the use of a switch with a total switching bandwidth of 102.4 Tbps, which can be configured as $64 \times 1.6$~Tbps, $128 \times 800$~Gbps, $256 \times 400$~Gbps, and $512 \times 200$~Gbps. Furthermore, we assume this switch is utilized to construct a medium-to-large-scale system comprising 65K NICs. The cost analysis for each topology is detailed in Table \ref{cost_ana}. 
\begin{table*}[t] 

  \caption{Cost-effectiveness comparison of various topologies when constructing a system with approximately 65K NICs. We assume an outbound bandwidth of 1.6~Tbps per NIC and that the entire network exclusively utilizes optical cables, without the use of any copper cables. The bare-metal price of a single 102.4~Tbps switch is assumed to be \$40,000, and the unit prices for 200~Gbps, 400~Gbps, 800~Gbps, and 1.6~Tbps optical transceivers are \$100, \$200, \$450, and \$1,200, respectively. Although the specific pricing information may deviate slightly from actual market conditions, the overall pricing trends and the reflected results should remain consistent. Notably, to maintain a reasonable bisection bandwidth, each switch in the second dimension of the MPHX(4, 86, 86, 9) network retains the same number of links as in the first dimension, \textit{i.e.}, 85. Consequently, within this second dimension, switches are interconnected by multiple links rather than a single link.}
  \label{cost_ana}
  \centering
  \begin{tabular}{c|c|c|c|c|c|c}
    \toprule 
    Topologies & $d$ &Switch configuration  & $N$ & $N_s$ & $N_o$ & Cost per NIC [\$]\\
    \midrule
    3-layer Fat-Tree & 4 & 64x1.6T & 65,536 & 5,120 & 393,126 (1.6T) & 10,323\\
    8-Plane 2-layer Fat-Tree & 2 & 512x200G & 65,536 & 3,072 & 2,097,152 (200G) & 5,075\\
    
    Dragonfly & 3 & 64x1.6T & 65,536 & 4,096 & 323,584 (1.6T) & 8,425\\
    Dragonfly+ & 3 & 64x1.6T & 65,536 & 4,096 & 327,680 (1.6T) & 8,500\\
    1-Plane 3D HyperX (MPHX(1,16,16,16,16)) & 3 & 64x1.6T & 65,536 & 4,096 & 315,392 (1.6T) & 8,275\\
    2-Plane 2D HyperX (MPHX(2,41,41,41))& 2 & 128x800G & 68,921 & 3,362 & 544,644 (800G) & 5,507\\
    4-Plane 2D HyperX (MPHX(4,86,86,9))& 2 & 256x400G & 66,564 & 3,096 & 1,058,832 (400G) & 5,041\\
    8-Plane 1D HyperX (MPHX(8,256,256))& 1 & 512x200G & 65,536 & 2,048 & 1,570,816 (200G) & 3,647\\
  \bottomrule
\end{tabular}
\end{table*}

As observed, constructing a three-tier Fat-Tree network with non-breakout switches incurs exorbitant costs. In contrast, as the number of network planes increases, the MPHX topology progressively demonstrates superior cost-effectiveness. Compared to the multi-plane Fat-Tree network, the average cost per NIC is reduced by 28.0\%. Furthermore, when factoring in the use of copper cables for connections between NICs and access switches, the cost-effectiveness of the MPHX network is further amplified.
\section{DISCUSSION}
\subsection{Other multi-plane topologies}

Theoretically, multi-plane technology can be applied to any topology. Let us first consider the multi-plane Dragonfly \cite{Dragonfly08} network. By employing multi-plane technology, switches can be broken out into a larger number of finer-grained ports, thereby reducing the network diameter. In the Dragonfly topology, each time the switch radix doubles, the number of global ports a router connects to also doubles, while the number of NICs per group quadruples, and the total number of network groups is reduced to a quarter. Consequently, as switch ports are further broken out, it becomes highly probable that the global ports of a single router will suffice to connect to all groups within the network. At this stage, the multi-plane Dragonfly network flattens into a multi-plane 2D HyperX network. Take the Frontier supercomputer as an example: its current switch radix is 64, each switch features 16 global ports, each group accommodates 512 NICs, and the network comprises 80 groups in total. If the switch ports are broken out into 128 finer-grained ports, each group will house 2,048 NICs, leaving the network with only 20 groups. Concurrently, each switch will possess 32 global ports, signifying that every switch can directly connect to all the remaining groups, thereby effectively forming a 2D HyperX network.

For the Dragonfly+ \cite{dragonfly+17} network, the conclusion is analogous to that of the Dragonfly topology. When increasing the switch radix, a single switch becomes highly likely to directly connect to all groups, thereby forming an architecture similar to a two-tier Fat-Tree combined with HyperX. As the switch radix is increased further, the network will likely be reduced to a single group and flatten into a multi-plane Fat-Tree network.

Regarding the recently proposed Zettafly-3 and Zettafly-4 networks \cite{zettafly25}, the findings parallel those of the Dragonfly and Dragonfly+ architectures. Increasing the switch radix eliminates the necessity for global switches, ultimately causing them to flatten into multi-plane HyperX and multi-plane Fat-Tree networks as well.

\subsection{Challenges in Implementing Multi-Plane Technology}
Despite its numerous advantages, deploying multi-plane technology is not trivial and imposes significant challenges on Network Interface Card (NIC) design. First, the NIC must be capable of uniformly distributing communication traffic across all planes, which essentially requires it to possess certain switching functionalities and the capability for out-of-order packet reception. Furthermore, for non-Fat-Tree multi-plane architectures, such as the multi-plane HyperX network proposed in this paper, it is additionally required that switches support adaptive routing. This is because the number of links between adjacent switches within a single plane is limited; consequently, the bandwidth of minimal paths is relatively low during cross-switch communication, necessitating the use of non-minimal paths to achieve effective load balancing.

\section{FUTURE WORK}

This paper provides only a preliminary overview of the multi-plane HyperX network. We plan to update this manuscript shortly to refine the technical details, propose a routing design for the multi-plane HyperX architecture, and present a comprehensive performance evaluation comparing it against topologies such as Dragonfly, Dragonfly+, multi-plane Fat-Tree, and Zettafly under synthetic traffic, as well as HPC and AI application workloads. We anticipate demonstrating the low-latency advantages of MPHX stemming from its reduced network diameter, and further highlighting its cost-effectiveness.
\bibliographystyle{ACM-Reference-Format}
\bibliography{Reference.bib}

\end{document}